\documentstyle[12pt,aaspp4]{article}
\begin{document}

\title{\bf EVOLUTION OF INTERSTELLAR CLOUDS IN
LOCAL GROUP DWARF SPHEROIDAL GALAXIES IN THE CONTEXT OF
THEIR STAR FORMATION HISTORIES}
\author{\bf HIROYUKI HIRASHITA$^{1,2}$}
\affil
{$^1$:  Department of Astronomy, Faculty of Science, Kyoto University,
Sakyo-ku, Kyoto 606-8502, Japan}
\affil
{$^2$:  Research Fellow of the Japan Society for the Promotion of
Science}
\centerline{Jan.~27th, 1999}
\centerline{email: hirasita@kusastro.kyoto-u.ac.jp}
\authoremail{hirasita@kusastro.kyoto-u.ac.jp}
\begin{abstract}

We consider evolution of interstellar clouds in Local Group
dwarf spheroidal galaxies (dSphs) in the context of their
observed star formation histories. The
Local Group dSphs generally experienced initial bursts of star
formations in their formation epochs ($\sim 15$ Gyr ago), when
hot gas originating from the supernovae can make
the cold interstellar clouds evaporate. We find that the
maximum size of evaporating cloud is 10 pc.
Thus, clouds larger than 10 pc can survive during the
initial star formation. These surviving clouds can
contribute to the second star formation to produce
``intermediate-age ($\sim 3$--10 Gyr ago) stellar populations.''
Assuming that
collisions between clouds induce star formation and
that the timescale of the second star formation is a few Gyr,
we estimate the total mass of the clouds. The total mass is
about $10^{4}M_\odot$, which is 1--3 orders of magnitude
smaller than the typical stellar mass of a present dSph.
This implies that the initial star formation is dominant over
the second star formation, which is broadly consistent with
the observed star formation histories. However, the variety of the
dSphs in their star formation histories suggests that
the effects of environments on the dSphs may be important.

\end{abstract}


\section{INTRODUCTION}

Recent observations have been revealing the properties of
the Local Group dwarf spheroidal galaxies (dSphs). The
dSphs have luminosities of order $10^{5\mbox{--}7}L_\odot$
and are characterized by their low surface brightnesses
(see Gallagher \& Wyse 1994 for review).

The dSphs contain such small amounts of gas that they show
no evidence of present star formation.\footnote{In fact, a dwarf
galaxy with present star formation is not generally called dSph.}
Saito (1979) showed
that instantaneous gas ejection from supernovae (SNe) can make the gas
in proto-dSphs escape in their initial burst of star formation
(see also Larson 1974).
This so-called SN feedback mechanism nicely accounts for the
observed scaling relations among mass, luminosity, and metallicity
of each dSph
(Dekel \& Silk 1986; see also Hirashita, Takeuchi, \& Tamura 1998).

The stellar population analyses of the Local Group dSphs
show that their star formation
histories are full of variety (Mateo et al. 1998; Aparicio 1999;
Grebel 1999). Some dSphs has prominent
 intermediate-age ($\sim 3$--10 Gyr ago) stellar populations,
and others have only small numbers of such populations.
A dSph located close to the Galaxy tends to have poor
intermediate-age stars. This may indicate that the Galaxy
has affected their star formation histories through
ultraviolet (UV) radiation or the Galactic wind (van den Bergh 1994).

In this paper, we aim at understanding of the evolution
of interstellar medium in the Local Group dSphs in the
context of their star formation histories.
Though it is generally difficult to infer the physical properties of 
interstellar gas of the dSphs in their star formation epochs,
star formation histories derived from the stellar color-magnitude
diagrams (e.g., Gallagher \& Wyse 1994)
help us to obtain some information on the physical quantities of
the gas. A merit of using the Local Group dSphs
is that they are so close to the Galaxy that their star formation
histories are directly inferred from their stellar populations.
This work can be applied to dwarf irregular
galaxies, elliptical galaxies or distant galaxies in the
future. 

This paper is organized as follows.
First of all, in the next section,
we consider the evolution of interstellar clouds in initial
star formation epoch.
Then, in \S 3, we apply the cloud-cloud collision model
to star formation in the intermediate ages of dSphs.
Finally, the last section is devoted to discussions.

\section{SURVIVAL OF CLOUDS IN PROTO-DWARF GALAXIES}

The collapse of gas in a proto-dSph induces the
initial burst of star formation. This initial burst occurs in
the dynamical timescale determined by the dark matter potential
($\sim 10^{7}$ yr). During the burst,
the hot gas (temperature of $T\sim 10^6$ K) originating mainly from
SNe (McKee \& Ostriker 1977)
contributes to heating of interstellar gas
through thermal conduction (e.g., Draine \& Giuliani 1984).
In this section, we examine the effect of the thermal conduction.

\subsection{Evaporation Timescale of Clouds}

It is widely accepted that
interstellar medium is a cloudy fluid (e.g., Elmegreen 1991).
Interstellar medium is multiphase gas with various temperature
and number density (McKee \& Ostriker 1977).
Here, we simply consider two-phase interstellar gas composed of
hot ($T\sim 10^6$ K) diffuse gas and cool ($T<10^4$ K) 
clouds. For the evolution of interstellar medium in the context of
multiphase interstellar medium, see e.g., Fujita, Fukumoto, \&
Okoshi (1996).

The hot gas originating from the successive SNe
can heat cool interstellar clouds.
The cool gas may finally evaporate. The evaporating gas as well as
the hot gas
escapes freely out of the proto-dwarf galaxy, since the thermal
energy at $T\sim 10^6$ K is much larger than 
the gravitational potential. The timescale for the gas to
escape out of the galaxy is estimated by crossing time
 $t_{\rm cross}$ defined by
\begin{eqnarray}
t_{\rm cross}\equiv \frac{R_{\rm dSph}}{c_{\rm s}},
\end{eqnarray}
where $c_{\rm s}$ is the sound speed of the hot gas (typically
100 km s$^{-1}$) and
$R_{\rm dSph}$ is the typical size of the proto-dSph.
Thus, the crossing time is estimated by
\begin{eqnarray}
t_{\rm cross}\simeq 1.0\times 10^7\left(
\frac{R_{\rm dSph}}{1~{\rm kpc}}\right)\left(
\frac{c_{\rm s}}{100~{\rm km}~{\rm s}^{-1}}\right)^{-1}~{\rm yr}.
\label{crtime}
\end{eqnarray}
 Hirashita, Takeuchi,
\& Tamura (1998) also estimated the crossing time.
The estimation from the view point of thermal energy
also provides the timescale of $10^{6\mbox{--}7}$ yr 
for the escape of hot gas from dwarf galaxies
(Yoshii \& Arimoto 1987; Nath \& Chiba 1995).

Now we discuss the process of evaporation.
After treating the conservation of mass and energy,
Cowie \& McKee (1977) derived the typical
mass loss rate ($\dot{m}$) of a cool cloud embedded in the hot
medium as
\begin{eqnarray}
\dot{m}=
\frac{16\pi\mu m_{\rm H}\kappa_{\rm h}R_{\rm c}}{25k_{\rm B}},
\end{eqnarray}
where $\mu$ is the mean weight of a particle normalized by the mass
of a hydrogen atom ($m_{\rm H}$),
$R_{\rm c}$ is the radius of the cloud
[a spherical cool ($T\ll T_{\rm h}$, where  $T_{\rm h}$ is the
temperature of the hot gas) cloud is
assumed], $k_{\rm B}$ is the Boltzmann constant, and
$\kappa_{\rm h}$ is the thermal conductivity estimated at the
temperature of $T_{\rm h}$ and the electron number density of
$n_{\rm h}$ (the electron number density of the hot gas).
The thermal conductivity is expressed as
\begin{eqnarray}
\kappa_{\rm h}=1.8\times 10^{-5}
\frac{T_{\rm h}^{5/2}}{\ln\Lambda}~{\rm
erg~s^{-1}~{deg}^{-1}~{\rm cm}^{-1}},
\end{eqnarray}
where $\ln\Lambda$ is the Coulomb logarithm which is a function of
electron number density and electron temperature
(Spitzer 1956; Cowie \& McKee 1977).
Using $\dot{m}$, the timescale for
the evaporation, $t_{\rm evap}$, of the cloud is estimated as
\begin{eqnarray}
t_{\rm evap}=\frac{m_{\rm c}}{\dot{m}}=1.4\times 10^7\left(
\frac{\bar{n}_{\rm c}}{1~{\rm cm}^{-3}}\right)\left(
\frac{R_{\rm c}}{10~{\rm pc}}\right)^2\left(
\frac{T_{\rm h}}{10^6~{\rm K}}\right)^{-5/2}\left(
\frac{\ln\Lambda}{30}\right)~{\rm yr},
\label{evtime}
\end{eqnarray}
where $\bar{n}_{\rm c}$ is the mean number density of gas in 
the cloud and
$m_{\rm c}$ is the mass of the cloud estimated by
\begin{eqnarray}
m_{\rm c}=\frac{4\pi}{3}R_{\rm c}^3\mu m_{\rm H}\bar{n}_{\rm c}.
\label{clmass}
\end{eqnarray}

According to Larson (1974), hot gas fills more than half
of the dwarf galaxy in a short timescale ($<10^6$ yr) if
successive and multiple SNe are considered.
Thus, our picture of continuous evaporation during the crossing
time is justified. However, we should note that the timescale
is largely dependent on stellar initial mass function.

\subsection{Cooling Timescale of the Hot Gas}

In this subsection, we estimate the cooling timescale $t_{\rm cool}$.
The cooling timescale is expressed by
\begin{eqnarray}
t_{\rm cool}=
\frac{3k_{\rm B}T_{\rm h}}{2n_{\rm h}\Lambda_{\rm cool} (T_{\rm h})},
\end{eqnarray}
where $n_{\rm h}$ is the number density of electrons in the hot gas
and $\Lambda_{\rm cool} (T_{\rm h})$ is the cooling function as
a function of temperature. The cooling function is composed of
two contributions as
\begin{eqnarray}
\Lambda_{\rm cool} (T_{\rm h})=\Lambda_{\rm ff} (T_{\rm h})+
\Lambda_{\rm line} (T_{\rm h}),
\end{eqnarray}
where $\Lambda_{\rm ff}$ and $\Lambda_{\rm line}$ represent
the cooling rates through free-free radiation and through
metal-line emission, respectively. According to
Gaetz \& Salpeter (1983), the metal cooling is estimated as 
\begin{eqnarray}
\Lambda_{\rm line} (T_{\rm h}=10^6~{\rm K})\simeq 1.3\times
10^{-22}\zeta~{\rm erg}~{\rm s}^{-1}~{\rm cm}^{3},
\end{eqnarray}
 where $\zeta$ is the metallicity normalized by the solar system
abundance (see also Raymond, Cox, \& Smith 1976 for the cooling
function).  On the other hand, the free-free cooling function is
estimated as
\begin{eqnarray}
\Lambda_{\rm ff} (T_{\rm h})\simeq 2\times 10^{-24}\left(
\frac{T_{\rm h}}{10^6~{\rm K}}
\right)^{1/2}~{\rm erg}~{\rm s}^{-1}~{\rm cm}^{3}
\end{eqnarray}
(Rybicki \& Lightman 1979).

If we assume $\zeta\sim 0.01$, which is a typical
value for stellar metallicity observed in the present dSphs
(Aaronson \& Mould 1985; Buonanno et al. 1985),
$\Lambda_{\rm line} (T_{\rm h}=10^6~{\rm K})\simeq 1.3\times
10^{-24}~{\rm erg}~{\rm s}^{-1}~{\rm cm}^{3}$. The resulting
cooling timescale becomes
\begin{eqnarray}
t_{\rm cool}\simeq 2\left(\frac{T_{\rm h}}{10^6~{\rm K}}\right)
\left(\frac{n_{\rm h}}{10^{-3}~{\rm cm}^{-3}}\right)^{-1}
\left(\frac{\Lambda_{\rm cool}}{3\times
10^{-24}~{\rm erg~s}^{-1}~{\rm cm}^{-1}}
\right)^{-1}~{\rm Gyr}.
\end{eqnarray}
Hence, the cooling timescale is much longer than the
crossing timescale (eq. \ref{crtime}). This means that the effect of
cooling of the hot gas can be neglected. The present stellar 
metallicity of the dSph sample is at most $\zeta\sim 0.1$
if we consider the error range. Adopting this upper limit value,
we obtain the shorter value of the cooling time of
$\sim 0.5$ Gyr. Even in this case, the cooling time is much shorter
the crossing timescale.\footnote{We note that we should consider the
metallicity of {\it gas}, not stars. Though the lack of gas
content in dSphs makes it impossible to know the metallicity of
their gas, we imagine that the gas metallicity did not exceed
0.1 from the present gas content in low-mass dwarf irregular
galaxies.}

We have considered uniform hot gas distribution.
The confinement of the hot gas might be possible. With the
confinement, the density of the hot gas can be so high that the
cooling timescale becomes shorter than the crossing timescale.
However, since the crossing time of the
confined region also becomes shorter than that in the previous
estimation, it seems difficult to realize the shorter cooling
timescale than the crossing time. Once the hot gas crosses a
dense region, the hot gas blows away easily. Thus, it is reasonable
to assume the longer cooling time than the crossing  time.
We note that if a physically reasonable confining mechanism of the
hot gas
is found the estimation in this section may need modification.

The mixing of the hot gas with the warm gas can make the
cooling timescale short.
The mixing produces the gas with
temperature of $\sim 10^5$ K (Begelman \& Fabian 1990;
Slavin, Shull, \& Begelman 1993), at which the line cooling rate
becomes an order of magnitude larger than that at  $\sim 10^6$ K
(Gaetz \& Salpeter 1983).
Thus, the cooling timescale may be shorter by an order of magnitude
than the previous estimate. However, even in this case,
the cooling time is larger than the crossing time.

\subsection{Condition for Survival of a Cloud}

Here, we estimate the size of a cloud that survives the
evaporation. Since the evaporation process is effective
in the timescale of $t_{\rm cross}$,
the condition for the survival is expressed by
\begin{eqnarray}
t_{\rm evap}>t_{\rm cross}.
\end{eqnarray}
From equations (\ref{crtime}) and  (\ref{evtime}), the above
condition is written as
\begin{eqnarray}
R_{\rm c} & > & R_{\rm crit}\nonumber \\ 
  & \equiv & 8
\left(\frac{\bar{n}_{\rm c}}{1~{\rm cm}^{-3}}\right)^{-1/2}
\left(\frac{T_{\rm h}}{10^6~{\rm K}}\right)^{5/4}
\left(\frac{\ln\Lambda}{30}\right)^{-1/2}\nonumber \\
& & \times
\left(\frac{R_{\rm dSph}}{1~{\rm kpc}}\right)^{1/2}
\left(\frac{c_{\rm s}}{100~{\rm km}~{\rm s}^{-1}}
\right)^{-1/2}~{\rm pc},
\end{eqnarray}
where we define the critical radius for the survival, $R_{\rm crit}$.
Thus, the cloud larger than $R_{\rm crit}\sim 10$ pc can survive
the thermal conduction during the initial burst of star formation.

The motions of interstellar clouds produce velocity shear between
the clouds and ambient hot gas. The shear may lead to the
Kelvin-Helmholtz (K-H) instability (Chandrasekhar 1961). Since the
growth timescale of
the K-H instability is as short as the conduction timescale
(Appendix A) for $R_{\rm c}\sim 10$ pc (for the notation in
Appendix A, $R_{\rm c}=\lambda$), the K-H instability
as well as the conduction can determine the minimum mass of the
clouds.

\section{CLOUD-CLOUD COLLISION}

One of the direct test of survival of the cloud in the initial
star formation in proto-dSph
is to examine star formation histories of the dSphs. Their star
formation histories are investigated from the stellar population
analyses (Gallagher \& Wyse 1994; Mateo 1998). The second star
formation produced what is called ``intermediate-age stellar
populations'' (e.g., Gallagher \& Wyse 1994). There are evidences
of second star formation in each Local Group dSphs.
Assuming that the second star formation is due
to the surviving clouds within a dSph, we examine the
physical properties of the clouds.

Before the examination,
we should fix the mechanism of the second star formation.
We assume that star formation is induced by collisions between
clouds. This physically means that the compression of
clouds during the collision makes free-fall time and cooling time
of the clouds shorter  (the physical process is described in
Kumai, Basu, \& Fujimoto 1993), which leads to formation of 
dense molecular clouds and finally to star formation.
Even if a cloud has formed stars, the shell of a stellar wind or a
supernova remnant associating with the cloud induce star formation
of another cloud which collides with it (see also
Roy \& Kunth 1995).  The idea that cloud-cloud collisions
induce star formation has a long history
(e.g., Field \& Saslaw 1965).
We note that another possible mechanism of star formation
should be investigated in future works.

The cloud-cloud collision timescale $t_{\rm coll}$ can be estimated
by
\begin{eqnarray}
t_{\rm coll}\simeq\frac{1}{N\sigma V},
\end{eqnarray}
where $N$ is the number of clouds per unit volume, $\sigma$ is
the geometrical cross section of a cloud, expressed as
$\sigma =\pi  R_{\rm c}^2$, and $V$ is the velocity of a cloud.
In fact, $N\sigma V$ should be written as $\langle N\sigma V\rangle$,
where $\langle\cdot\rangle$ means that the physical quantity is 
averaged for all the clouds. The number of the clouds is
the largest in the region where the gravitational potential well
is the deepest. The size of the deepest region is typically
estimated by the core radius (typically 100--1000 pc for dSphs;
e.g., Mateo 1998). Thus, the physical quantities in this 
section represents the typical values within the core radius.

The velocity $V$ is determined by the virial-equilibrium
value in gravitational potential of a dSph
($\sim 10$ km s$^{-1}$). If
we give a typical collision
timescale, we can estimate the number density of clouds in a dSph
by
\begin{eqnarray}
N\simeq\frac{1}{\pi R_{\rm c}^2Vt_{\rm coll}}.\label{clnumber}
\end{eqnarray}

Observationally, the duration of second star formation 
seems a few Gyr (van den Bergh 1994; Mateo et al. 1998; Grebel 1999).
Based on the assumption that the timescale of star formation
is determined by the collision timescale of interstellar clouds,
$t_{\rm coll}$ should be $\sim\mbox{a few Gyr}$. Thus, the following
estimation for $N$ is possible according to equation (\ref{clnumber}):
\begin{eqnarray}
N\simeq 1.0\times 10^{-7}\left(\frac{R_{\rm c}}{10~{\rm pc}}\right)^{-2}
\left(\frac{V}{10~{\rm km}~{\rm s}^{-1}}\right)^{-1}\left(
  \frac{t_{\rm coll}}{3~{\rm Gyr}}\right)^{-1}~{\rm pc}^{-3},
\label{density}
\end{eqnarray}
where we estimated the size of the cloud by $R_{\rm crit}$ estimated
in \S 2.3. This value of $N$ corresponds to the mean cloud-cloud
interval of $\sim 200$ pc. We note that $R_{\rm crit}$ is the lower
limit of the size of a cloud which survived the first star formation.
If the typical size is larger, $N$ becomes smaller and the typical
interval between clouds gets larger. 

This argument about the cloud size becomes clear if we introduce the
mass spectrum of clouds. The mass spectrum ${\cal N}(m)dm$ is defined
by the number density of clouds of mass between $m$ and $m+dm$.
If we assume that the spectrum can be expressed by the power-law,
${\cal N}(m)\propto m^{-p}$, and that mass ($m$) and size ($R$) of
any cloud is related by $m\propto R^3$ (the constant mass density
of any cloud), we obtain the size distribution of the clouds:
${\cal N}(R)dR\propto R^{-3p+2}dR$. Using this form, we calculate
the mean size of the clouds, $\langle R\rangle$, as
\begin{eqnarray}
\langle R\rangle & = & \int^{R_{\rm up}}_{R_{\rm crit}}R{\cal N}
(R)dR\left/\int^{R_{\rm up}}_{R_{\rm crit}}{\cal N}(R)dR\right.
\nonumber \\
& = & \left(\frac{3p-4}{3p-3}\right)
\left[\frac{(R_{\rm up}/R_{\rm crit})^{-3p+4}-1}
{(R_{\rm up}/R_{\rm crit})^{-3p+3}-1}\right] R_{\rm crit} ,
\end{eqnarray}
where $R_{\rm up}$ is the upper cutoff of the size.
Setting as $R_{\rm up}\gg R_{\rm crit}$, we obtain
$\langle R\rangle\sim R_{\rm crit}$ for $p>1$. In this case, 
estimation of $R_{\rm c}$ with $R_{\rm crit}$ (eq. \ref{density}) is
justified. Field \& Saslaw (1965) suggested $p=3/2$ to explain
observed star formation rate.

We further estimate the total mass of surviving clouds.
The total mass $M_{\rm c}$ can be estimated by
\begin{eqnarray}
M_{\rm c}& \simeq & \frac{4\pi}{3}R_{\rm dSph}^3Nm_{\rm c}\nonumber \\
& \simeq &
5\times 10^{4}\left(\frac{{R}_{\rm dSph}}{1~{\rm kpc}}
\right)^3\left(\frac{\bar{n}_{\rm c}}{1~{\rm cm}^{-3}}
\right)\left(\frac{R_{\rm c}}{10~{\rm pc}}\right)
\left(\frac{V}{10~{\rm km}~{\rm s}^{-1}}\right)^{-1}\left(
\frac{t_{\rm coll}}{3~{\rm Gyr}}\right)^{-1}M_\odot ,
\end{eqnarray}
where $m_{\rm c}$ is defined in equation (\ref{clmass}).
Thus, the total mass of the stellar population, which is formed
in the second star formation epoch (so-called intermediate age),
is roughly $\sim 10^4M_\odot$ (or less, since we can hardly expect all
the gas is converted to stellar mass). This is 1--3 orders of
magnitude smaller than the typical
stellar mass of the dSphs ($\sim 10^{5\mbox{--}7}M_{\odot}$).
Indeed, the number of the intermediate-age population is much
smaller than the old stellar population for the dSph, though
the Carina dSph has prominent intermediate-age populations.
(e.g., Mateo 1998). We will comment
on the exception later in \S 4.

Here, we should mention the effect of SNe
during the second
star formation. As shown above, the second star formation
is not so active as the initial burst of star formation.
Thus, once cloud size is determined during the initial 
burst through the thermal conduction, the SN heating in the
intermediate age has little influence on the clouds.
Thus, we can reasonably ignore the SN heating in the
intermediate age.

\section{DISCUSSIONS AND IMPLICATIONS}

We have inferred the evolution of interstellar clouds in 
the Local Group dSphs from their observed star
formation histories.
Owing to the hot gas supplied by initial star formation,
small interstellar clouds evaporates. However, clouds larger than
$\sim 10$ pc can survive during the burst of star formation.
The surviving clouds contribute to second star formation to
form so called ``intermediate-age stellar populations.''

There are observational evidences that the second star formation
occurred in the intermediate age ($\sim3$--10 Gyr ago). The timescale of 
the second star formation is typically a few Gyr
(e.g., Mateo et al. 1998). Assuming that
star formations are induced by cloud-cloud collisions, 
the collision timescale should be a few Gyr to realize the
observed timescale of the second star formation.
Since the collision timescale is related to the number of clouds,
the number is constrained. The expected number density of clouds is
typically $1.0\times 10^{-7}$ pc$^{-3}$, which indicates that
the total mass of gas contributing to the second
star formation is typically $10^4M_\odot$. This is 1--3
orders of magnitude smaller than the observed stellar mass
of the Local Group dSphs. This indicates that almost all the
stars in the dSphs are formed in the initial star formation.

Recently, Hirashita, Takeuchi, \& Tamura (1998) suggested that
the luminous mass of a dSph is determined by the depth of
dark matter potential. Their suggestion is true if the first star
formation is dominant in the star formation histories of dSphs.
Indeed, our cloud-cloud collision model implies that
the second stellar population is not dominant in mass.

We should also consider environmental effects.
Van den Bergh (1994) suggested that
environmental effects on dSphs may be important for
their star formation histories. He showed that the star formation
histories of the Local Group dwarf galaxies correlate with the
Galactocentric distances
(the distances from the Galaxy): Dwarf galaxies near the Galaxies,
such as Ursa Minor and Draco contain only a little fraction of
intermediate-age or recent stellar population, while
there is observational evidences of recent star formations
in distant dwarf galaxies. In the same paper, he also suggested that
star formations in the dwarf galaxies are affected
by the existence of UV radiation or the wind from the
Galaxy. Hirashita, Kamaya, \& Mineshige (1997) showed
that the Galactic wind can strip the gas of nearby
dwarf galaxies. In the epoch of initial burst, the effects of
OB star radiations and SNe in dSphs are much stronger than 
environmental effects and determine the structure of dSphs
(Hirashita, Takeuchi, \& Tamura 1998). However, after the first
star formation, such effects become weak, so that the dominant
factor to determine the physical condition is environmental
effects such as the Galactic wind (Hirashita, Kamaya, \& Mineshige
1997). Thus, the environmental effect may be responsible for 
the physical nature of the second star formation.
Since Ursa Minor and Draco are located closer
to the Galaxy than other dSphs, they are easily affected
by the ram pressure of Galactic wind which strips the gas of them.
Therefore, the second star formation is less prominent in these two
dSphs than other dSphs (van den Bergh 1994). Recently, the star
formation histories of the companion dSphs of M31 has begun to
be made clear (e.g., Armandroff, Davies, \& Jacoby 1998).
The increase of the number of sample dSphs will contribute to test 
of the environmental effects.
Here we should note that Einasto, Saar, \& Kaasik (1974) pointed
out environmental effects on structures of galaxies.

Another environmental effect is possible. The burst of star
formation may be induced by infall of intergalactic gas clouds.
If a cloud is captured in a gravitational potential well of
a dSph, the cloud may form stars. Hirashita, Kamaya, \& Mineshige
(1997) pointed out that the present activity of star formation
in the Magellanic Clouds may be due to such infall of gas.
However, since potential wells of dSphs are much shallower than
those of the Magellanic Clouds, it seems difficult for dSphs to
capture the intergalactic clouds.

Contrary to most dSphs, 
the Carina dSph shows a burst of star formation in the intermediate
age. One possibility to explain this peculiar nature of the galaxy
is UV radiation field. It is suggested that UV radiation field
suppresses formation of dwarf galaxies
(Babul \& Rees 1992; Efstathiou 1992).
The UV from the Galaxy may have suppressed the formation of
Carina. When the UV field at the galaxy became weak in the
intermediate age, Carina experienced a burst of star formation.
To show that this is true, we also explain why most of
dSphs did not suffer the suppression from UV radiation.
The future work on the history of UV radiation field in the
Local Group may provide us a hint to solve this problem.

We note that the cloud-cloud collision model is applicable
to the star formation histories of dwarf irregular
galaxies, giant elliptical galaxies or distant galaxies.
For the application of the collision model to dense molecular clouds
in a high-redshift object, see e.g., Ohta et al. (1998).

The following points remain to
be solved in this paper:

[1] What determines the number of stars formed in the initial burst
of star formation? This is the problem raised also in Hirashita, 
Takeuchi, \& Tamura (1998).

[2] What determines the number of clouds that contribute to
the intermediate-age stellar populations?
This question is related to the mass spectrum of
interstellar clouds (\S 3).

\acknowledgements

We wish to thank the anonymous referee for helpful comments
that substantially improved the discussion of the paper.
We are grateful to S. Mineshige for continuous encouragement. 
We thank H. Kamaya,
T.~T.~Takeuchi, H.~Nomura,  N.~Tamura, and K.~Yoshikawa for
helpful comments
and useful discussions. This work was motivated by discussions
with  S. van den Bergh at the IAU meeting. We would like to
thank him for his stimulating comments.
This work is supported by the Research Fellowship of the Japan
Society for the
Promotion of Science for Young Scientists. We fully utilized the
NASA's Astrophysics Data System Abstract Service (ADS).

\newpage

\appendix

\centerline{\bf APPENDIX}

\section{THE KELVIN-HELMHOLTZ INSTABILITY}

The Kelvin-Helmholtz (K-H) instability has been discussed in various
astrophysical contexts. For example, Miyahata \& Ikeuchi (1995)
discussed the stability of protogalactic cloud against the
K-H instability (see also Murray et al. 1993). 
The K-H instability is also applied to interstellar
physics: Klein, McKee, \& Collella (1994) examined the interstellar
clouds, while Fleck (1984) and Kamaya (1996) investigated
molecular clouds. All of these works assume that dense
clouds are embedded in a diffuse medium. Since the dense medium
generally moves at the velocity determined by the depth of
gravitational potential, there is generally a relative motion between
the dense and diffuse media. The presence of the relative motion
makes it possible to discuss the K-H instability
(e.g., Chandrasekhar 1961).

In \S 2, we considers interstellar clouds embedded
in  hot tenuous gas originating from successive SNe.
Since clouds in a galaxy generally move at the velocity 
determined by the gravitational potential of the galaxy,
it is necessary to examine the timescale of the growth of
the K-H instability.
The growth rate ($\omega$) in the linear regime at a flat interface
between the cloud and ambient hot gas can be expressed as
\begin{eqnarray}
\omega =k
\frac{(\rho_{\rm c}\rho_{\rm h})^{1/2}U}
{\rho_{\rm c}+\rho_{\rm h}},
\end{eqnarray}
where $k$ is a wavenumber of a mode ($k=2\pi /\lambda$, where
$\lambda$ is the wavelength), $\rho_{\rm c}$ and
$\rho_{\rm h}$ are, respectively, the mass densities of cloud
and hot gas, and $U$ is the relative velocity
(Drazin \& Reid 1981). Since
$\rho_{\rm c}\gg\rho_{\rm h}$, we obtain the following 
estimation for a typical proto-dwarf galaxy
considered in \S 2:
\begin{eqnarray}
\omega\simeq 6.4\times 10^{-14}\left(
\frac{\lambda}{10~{\rm pc}}\right)^{-1}\left(
\frac{\rho_{\rm h}/\rho_{\rm c}}{10^{-3}}\right)^{1/2}
\left(\frac{U}{10~{\rm km}~{\rm s}^{-1}}\right)~{\rm s}^{-1}.
\end{eqnarray}
The timescale of the growth, $t_{\rm KH}$, is estimated by
\begin{eqnarray}
t_{\rm KH}\equiv\frac{2\pi}{\omega}\simeq 3.1\times 10^7
\left(
\frac{\lambda}{10~{\rm pc}}\right)\left(
\frac{\rho_{\rm h}/\rho_{\rm c}}{10^{-3}}\right)^{-1/2}
\left(\frac{U}{10~{\rm km}~{\rm s}^{-1}}\right)^{-1}~{\rm yr},
\end{eqnarray}
which is comparable with the evaporation time defined in \S2.
Thus, the instability may determine the size of clouds
(see e.g., Kamaya 1998 for stabilizing effects).

\end{document}